\documentclass
[lengthestimate,tighten,
amssymb,aps,twocolumn,floatfix,tightenlines,superscriptaddress]{revtex4-1}
\pdfoutput=1
\usepackage{graphics}
\usepackage{graphicx,color}
\usepackage{subfigure}
\usepackage{float}
\usepackage[ansinew]{inputenc}
\usepackage{graphicx}
\usepackage{amsmath}
\usepackage{amssymb}
\usepackage{dsfont} 
\usepackage{hyphenat}
\usepackage[percent]{overpic}
\usepackage{xcolor}
\usepackage{appendix}
\usepackage{bbold}

\newcommand{\ket}[1]{\left | #1 \right \rangle}
\newcommand{\bra}[1]{\left \langle #1 \right |}
\newcommand{\proj}[1]{\ket{#1}\!\!\bra{#1}}

\DeclareMathOperator{\Tr}{Tr}
\newcommand{\argmin}{\operatornamewithlimits{argmin}}
\newcommand{\ketbra}[2]{\ket{#1}\!\!\bra{#2}}
   
\newcommand{\abs}[1]{\left | #1 \right |}

\begin{document}

\title{Universal Linear Optics}
\author{Jacques Carolan}
\affiliation{Centre for Quantum Photonics, H. H. Wills Physics Laboratory \& Department of Electrical and Electronic Engineering, University of Bristol, Merchant Venturers Building, Woodland Road, Bristol, BS8 1UB, UK}
\author{Chris Harrold}
\affiliation{Centre for Quantum Photonics, H. H. Wills Physics Laboratory \& Department of Electrical and Electronic Engineering, University of Bristol, Merchant Venturers Building, Woodland Road, Bristol, BS8 1UB, UK}
\author{Chris Sparrow}
\affiliation{Centre for Quantum Photonics, H. H. Wills Physics Laboratory \& Department of Electrical and Electronic Engineering, University of Bristol, Merchant Venturers Building, Woodland Road, Bristol, BS8 1UB, UK}
\affiliation{Department of Physics, Imperial College London, SW7 2AZ, UK}
\author{Enrique Mart\'{i}n-L\'{o}pez}
\affiliation{Centre for Quantum Photonics, H. H. Wills Physics Laboratory \& Department of Electrical and Electronic Engineering, University of Bristol, Merchant Venturers Building, Woodland Road, Bristol, BS8 1UB, UK}
\affiliation{Nokia Research Centre, Broers Building, 21 J.J. Thomson Avenue, Cambridge, CB3 0FA, UK}
\author{Nicholas J. Russell}
\author{Joshua W. Silverstone}
\affiliation{Centre for Quantum Photonics, H. H. Wills Physics Laboratory \& Department of Electrical and Electronic Engineering, University of Bristol, Merchant Venturers Building, Woodland Road, Bristol, BS8 1UB, UK}
\author{Peter J. Shadbolt}
\affiliation{Department of Physics, Imperial College London, SW7 2AZ, UK}
\author{Nobuyuki Matsuda}
\affiliation{NTT Basic Research Laboratories, NTT Corporation, 3-1 Morinosato-Wakamiya, Atsugi, Kanagawa 243-0198, Japan}
\author{Manabu Oguma}
\author{Mikitaka Itoh}
\affiliation{NTT Device Technology Laboratories, NTT Corporation, 3-1 Morinosato-Wakamiya, Atsugi, Kanagawa 243-0198, Japan}
\author{Graham D. Marshall}
\author{Mark G. Thompson}
\author{Jonathan C. F. Matthews}
\affiliation{Centre for Quantum Photonics, H. H. Wills Physics Laboratory \& Department of Electrical and Electronic Engineering, University of Bristol, Merchant Venturers Building, Woodland Road, Bristol, BS8 1UB, UK}
\author{Toshikazu Hashimoto}
\affiliation{NTT Device Technology Laboratories, NTT Corporation, 3-1 Morinosato-Wakamiya, Atsugi, Kanagawa 243-0198, Japan}
\author{Jeremy L. O'Brien}
\author{Anthony Laing}
\email{anthony.laing@bristol.ac.uk}
\affiliation{Centre for Quantum Photonics, H. H. Wills Physics Laboratory \& Department of Electrical and Electronic Engineering, University of Bristol, Merchant Venturers Building, Woodland Road, Bristol, BS8 1UB, UK}

\date{\today}

\begin{abstract}
\noindent
Linear optics underpins tests of fundamental quantum mechanics and computer science, as well as quantum technologies.
Here we experimentally demonstrate the longstanding goal of 
a single reprogrammable optical circuit that is sufficient to implement all possible linear optical protocols up to the size of that circuit.
Our six-mode universal system consists of a cascade of 15 Mach-Zehnder interferometers with 30 thermo-optic phase shifters integrated into a single photonic chip that is electrically and optically interfaced for arbitrary setting of all phase shifters, input of up to six photons and their measurement with a 12 single-photon detector system.
We programmed this system to implement heralded quantum logic and entangling gates, boson sampling with verification tests, and six-dimensional complex Hadamards.  We implemented 100 Haar random unitaries with average fidelity 0.999 $\pm$ 0.001.
Our system is capable of switching between these and any other linear optical protocol in seconds. 
These results point the way to applications across fundamental science and quantum technologies.  
\end{abstract}

\maketitle
\begin{figure*}[t!]
\includegraphics[width=1.0\linewidth]{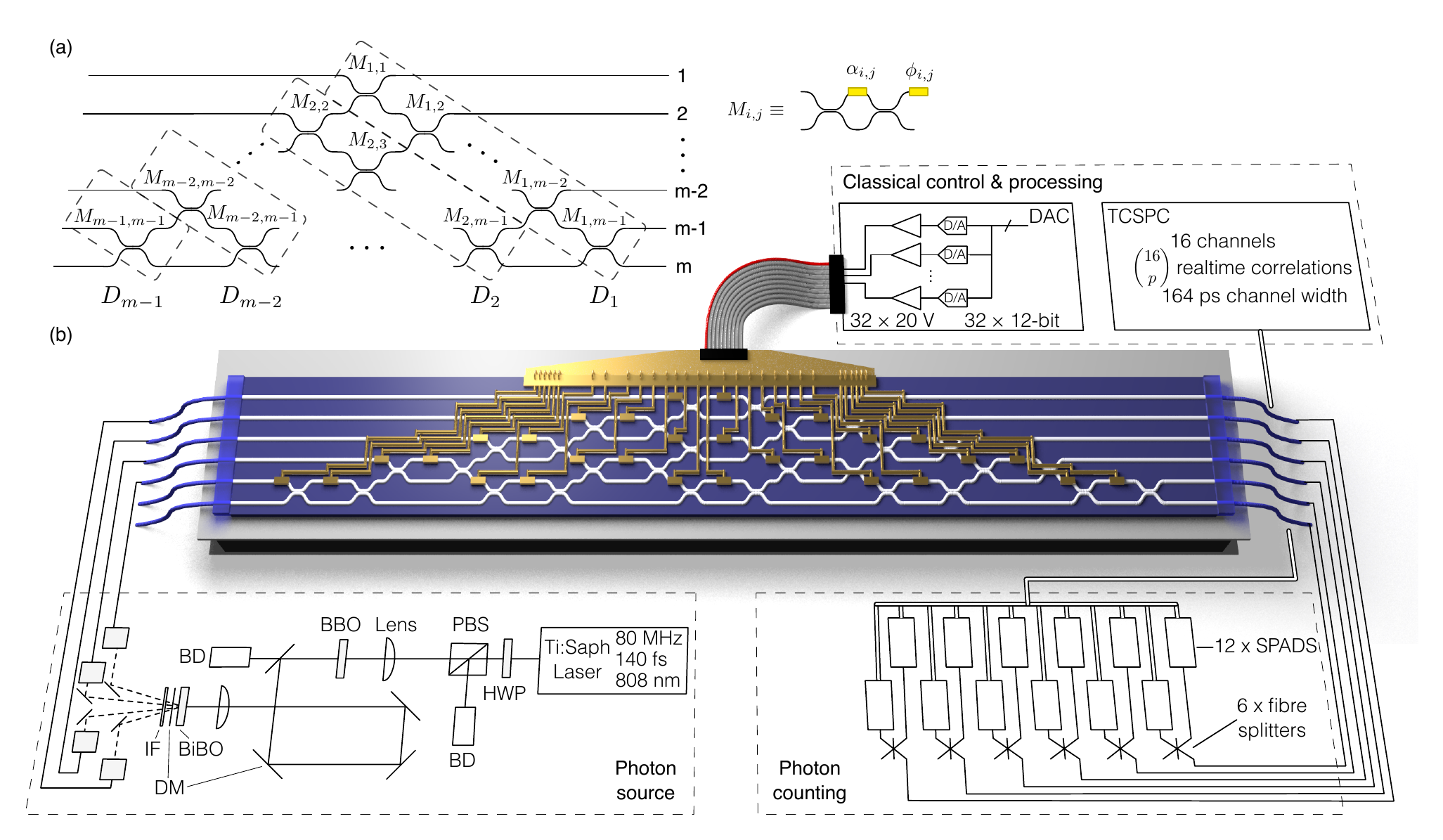}
\vspace{-.7cm}
\caption{
Universal linear optical processor (LPU).
(a)
Decomposition of a fully parametrised unitary for an $m$-mode circuit to realise any LO operation.  
Sub-unitaries $D_i$ consist of Mach-Zehnder interferometers $M_{i,j}$ (MZIs) built from phase shifters (yellow) and beam splitters, to control photon amplitudes ($\alpha_{i,j}$) and phases ($\phi_{i,j}$).
(b)
Multi-photon ensembles are generated via spontaneous parametric down-conversion (SPDC), comprising a BiBO crystal, dichroic mirrors (DM) and interference filter (IF); preceded by a pulsed Ti:sapphire laser and second harmonic generation from a BBO crystal.
Photons are collected into polarisation maintaining fibers and delivered to the LPU via a packaged v-groove fibre array.
The processor is constructed over six modes as a cascade of 15 Mach-Zehnder interferometers, controlled by 30 thermo-optic phase shifters, set by a digital-to-analgoue converter (DAC) and actively cooled by a Peltier cooling unit.
Photons are then out-coupled into a 2nd packaged VGA and sent to 6 (or 12 with fibre splitters for single-mode photon number resolving capability) single photon avalanche diodes (SPADs)
and counted using a 12-channel time-correlated single-photon counting module (TCSPC).  See Appendix for further details.
}
\vspace{-.3cm}
\label{fig:fig1}
\end{figure*}

Photonics has been crucial in establishing the foundations of quantum mechanics \cite{Pan:2012kv}, and more recently has pushed the vanguard of efforts in understanding new non-classical computational possibilities. Typical protocols involve nonlinear operations, such as the generation of quantum states of light through optical frequency conversion \cite{Ou:1992wj, Kwiat:1995ck}, or measurement-induced nonlinearities for quantum logic gates \cite{Knill:2001vi}, together with linear operations between optical modes to implement core processing functions \cite{Kok:2007ep}. Encoding qubits in the polarisation of photons has been particularly appealing for the ability to implement arbitrary linear operations on the two polarisation modes using a series of wave plates \cite{Langford:2007ug}. For path encoding the same operations can be mapped to a sequence of beamsplitters and phase shifters. In fact, since any linear optical (LO) circuit is described by a unitary operator, and a specific array of basic two-mode operations is mathematically sufficient to implement any unitary operator on optical modes \cite{Reck:1994dz}, it is theoretically possible to construct a single device with sufficient versatility to implement any possible LO operation up to the specified number of modes.

Here we report the realisation of this longstanding goal with a six-mode device that is completely reprogrammable and universal for LO.
We demonstrate the versatility of this universal LO processor (LPU) by applying it to several quantum information protocols, including tasks that were previously not possible.
We implement heralded quantum logic gates at the heart of the circuit model of LO quantum computing \cite{Knill:2001vi} and new heralded entangling gates that underpin the measurement-based model of LO quantum computing \cite{Raussendorf:2001vja, Nielsen:2004io, Browne:2005dd}, both of which are the first of their kind in integrated photonics.
We perform 100 different boson sampling \cite{Aaronson:2011tja,Broome:2013ti,Spring:2013to,Tillmann:2012ux,Crespi:2012fu} experiments and simultaneously realise new verification protocols.
Finally, we use multi-particle quantum interference to distinguish six-dimensional complex Hadamard operations, including newly discovered examples, where full classification remains an open mathematical problem. 
The results presented required reconfiguration of this single device to implement $\sim$1,000 experiments.

Isolated quantum mechanical processes, including lossless LO circuits acting on photons, preserve orthogonality between input and output states, and are therefore described by unitary operators.
While the relevant mathematics for parameterising unitary matrices has been known for at least a century  \cite{Hurwits:wh}, the theoretical formulation of a LO circuit in which an arbitrary unitary operator could be realised is more recent \cite{Reck:1994dz}.
The full space of $m$-dimensional unitary matrices (or unitaries) can be parametrised as a product of $\approx m^{2}/2$ two-dimensional unitary primitives, each with two free parameters.
Any given unitary then corresponds to a unique set of parameter values.
An arbitrarily reconfigurable LO network would be realised by implementing these unitary primitives as two-mode Mach-Zehnder interferometers (MZIs) with two beamsplitters (or integrated directional couplers) and two phase shifters, 
such that any given LO network corresponds to a set of phase shift 
values.

Such operation can be understood, with reference to the schematic shown in Fig.~\ref{fig:fig1}(a), as a sequence of $m$ sub-unitaries $D_{i}$, each of which is guaranteed to enable transformation of the state of a photon input in its top mode into an arbitrary superposition of all its output modes.
The top MZI in each $D_{i}$ is set to retain the desired probability for a photon occupying the top mode, before imparting the desired phase on the top mode.
The remainder of the photon undergoes a similar operation on the subsequent modes.
Full reconfigurability is realised by feeding all output modes from one $D_{i^{\prime}}$ into all but the uppermost input mode of the following $D_{i^{\prime}-1}$ \cite{Miller:2012gn}.

Realising such a scheme requires sub-wavelength stability and high fidelity components to support both classical and quantum interference --- possibilities opened up by integrated quantum photonics \cite{Politi:2008tl,Obrien:2009eu,Marshall:2009ub,Crespi:2011cy,Matthews:2009gi,Smith:2009hm,Laing:2010fk,Peruzzo:2010tq,Sansoni:2012eq}.
A schematic of our LPU is shown in Fig.~\ref{fig:fig1}(b).
The device, made with planar lightwave circuit (PLC) technology \cite{Himeno:1998gw,Takahashi:2011fk}, comprises an array of 30 silica-on-silicon waveguide directional couplers with 30 electronically controlled thermo-optic phase shifters, to form a cascade of 15 MZIs across six modes (see Appendix for further details).
Such an LPU can implement any LO protocol, and here we focus on several protocols at the forefront of quantum information science and technology.\\

\noindent\textbf{\textit{Quantum gates:}} 
With the addition of single photon sources and measurements, and rapid feed-forward of classical information, both the circuit \cite{Knill:2001vi} and measurement-based \cite{Raussendorf:2001vja, Nielsen:2004io, Browne:2005dd} models of digital quantum computing can be efficiently implemented with LO.
Basic two-qubit processes are realised probabilistically with LO circuitry, therefore a key requirement for scalability is that a successful operation is heralded by the detection of ancillary photons, to signal that the processed photonic qubits are available for use in the larger architecture \cite{Gasparoni:2004hp}.

We programmed our device to implement
a new compact four-photon scheme, suitable for measurement-based quantum computing, which generates the entangled state of two photonic qubits upon the detection of another two \emph{ancilla} photons, schematically displayed in
\onecolumngrid
\begin{center}
\begin{figure*}[h!]
\includegraphics[trim=0 0 0 0, clip, width=.93\linewidth]{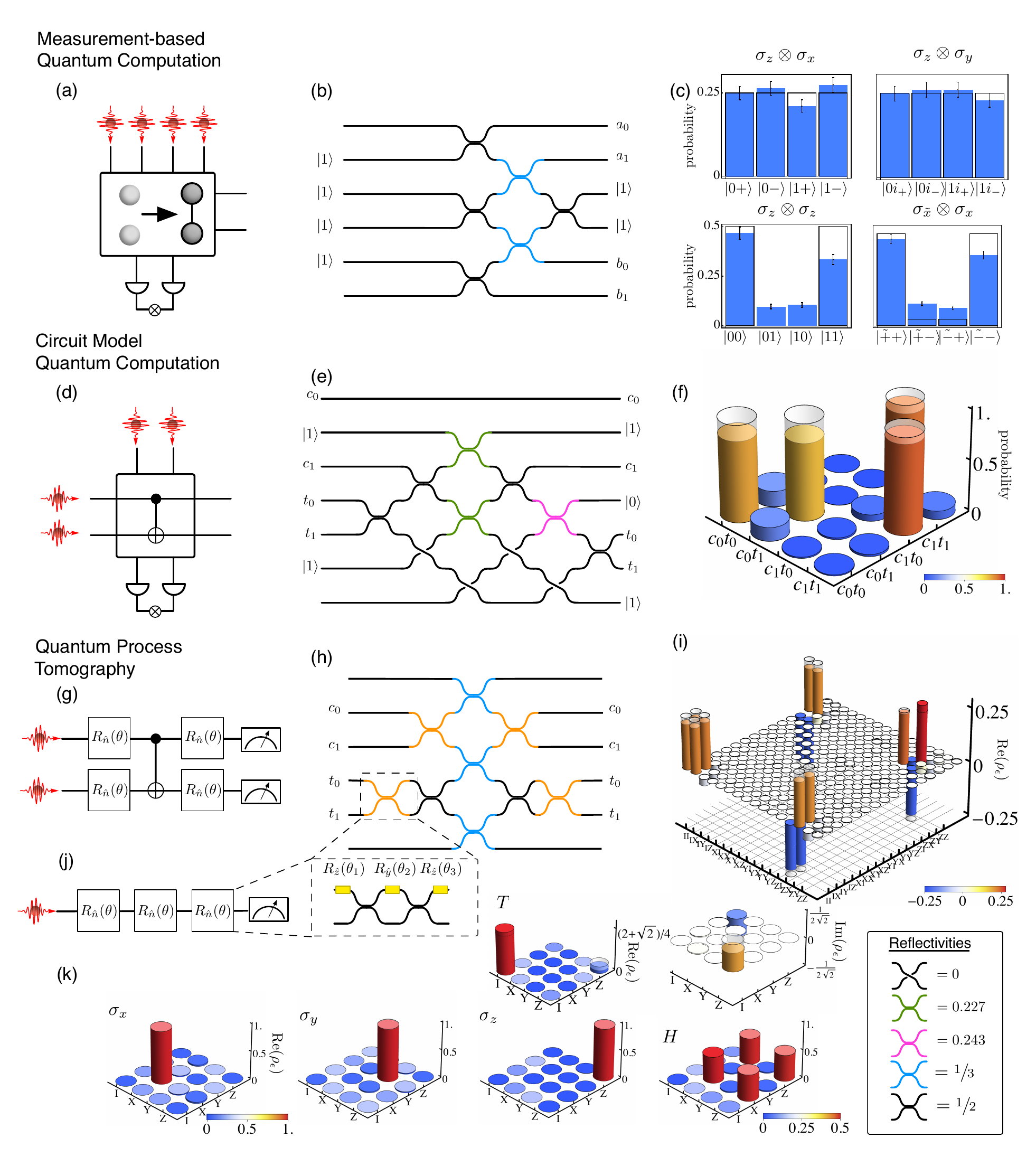}
\vspace{-.5cm}
\caption{Gates for linear optical quantum computing.
(a)
The heralded Bell state generator receives four photons and emits two of them in a maximally entangled state upon detection of the remaining two.
(b)
Our linear optical protocol emits a Bell state on modes $\{a,b\}$ with input and heralding modes labelled by $\ket{1}$.
(c)
Experimental data measuring photon correlations (blue), with error bars from Poissonian counting statistics.
The overlaid ideal theoretical values show that measurements in common bases should be correlated
while measurements in different bases should be uncorrelated (see Appendix for details).
(d)
The heralded CNOT gate is successfully implemented on the two photonic qubits, upon detection of the two ancilla photons.
(e)
The linear optical protocol realising the heralded CNOT operation on the control $c_{0,1}$ and target $t_{0,1}$ qubits.
(f)
Experimental data showing the computational truth table, with the ideal theoretical truth table overlaid.
(g)
Quantum process tomography of an unheralded (or post-selected) two-qubit CNOT gate can be performed with the addition of arbitrary single qubit preparation and measurement operations.
(h)
The linear optical circuit realising the post-selected CNOT gate, with MZIs (inset) allowing single qubit operations.
(i)
Experimentally determined process matrix with ideal theoretical overlaid.
(j)
An arbitrary single qubit operation can be realised with an MZI and additional phase shifters.  Three consecutive MZIs allows us to perform full process tomography on any single qubit operation.
(k)
Experimental data showing the measured process matrices for the three Pauli operations, the Hadamard gate ($\hat{H}$), and the $\pi/8$ phase gate ($\hat{T}$).  Experimental data is corrected for measured detector efficiencies.
}
\label{fig:fig2}
\end{figure*}
\end{center}
\twocolumngrid
\noindent  Fig.~\ref{fig:fig2}(a,b).
For Bell states, measurements in common bases should be correlated while measurements in different bases should be uncorrelated.
We implemented both types of measurement, data shown in Fig.~\ref{fig:fig2}(c), finding the mean statistical fidelity $\mathcal{F}_s=\sum_i \sqrt{p^{\text{exp}}_i.p^{\text{th}}_i}$ between all four experimental $p^{\text{exp}}$ and theoretical $p^{\text{th}}$ probability distributions to be $\mathcal{F}_s=0.966\pm0.004$ (see Methods for an explanation of all error analyses).
We used these measurements to verify the entanglement of our state by calculating
$E = 1/2 (\langle\sigma_x\otimes\sigma_x\rangle +  \langle\sigma_z\otimes\sigma_z\rangle )$,
finding a value of $E=0.673\pm0.031$ where $E > 1/2$ witnesses entanglement \cite{Wunderlich:2009dpa}.

Next, we reprogrammed the device to realise a quantum logic gate suitable for use in the circuit model of quantum computing [Figs.~\ref{fig:fig2}(d,e)].
The detection of two ancilla photons heralds the implementation of a controlled-NOT (CNOT) operation \cite{Knill:2001vi,Ralph:2001jc,Okamoto:2011ei} between two photonic qubits: the logical state of the \emph{target} qubit is flipped (0 $\leftrightarrow$ 1) if the \emph{control} qubit is in the state 1, and left unchanged if the control qubit is in state 0.  
We measured the logical truth table for 
this operation, shown in Fig.~\ref{fig:fig2}(f), and found its mean statistical fidelity averaged over all computational inputs to be $\mathcal{F}_s=0.927\pm0.004$ with the ideal.

While these are the first heralded two-qubit processes in integrated optics, several examples of photonic chips specifically fabricated to implement \emph{unheralded} (and hence not scalable) two-qubit gates have been reported \cite{Crespi:2011cy,Shadbolt:2011bw,Li:2013ie}.
To compare the performance our universal processor against devices fabricated for a specific task, we implemented an unheralded two-photon CNOT gate \cite{Ralph:2002id,Hofmann:2002hl,OBrien:2003kz} with single qubit preparation and measurement capabilities [Figs.~\ref{fig:fig2}(g,h)] and performed full quantum process tomography \cite{Chuang:1997jx} [Fig.~\ref{fig:fig2}(i)].  The process fidelity was found to be
$\mathcal{F}_p=0.909 \pm 0.001$, and the average gate fidelity 
$\mathcal{F}_g=0.927 \pm 0.001$ (see Appendix for details), greater than those previously reported \cite{OBrien:2004cn,Crespi:2011cy,Shadbolt:2011bw}.

Combined with two-qubit operations, a small set of single qubit gates [Fig.~\ref{fig:fig2}(j)], including the Hadamard ($\hat{H}$) and $\pi/8$ ($\hat{T}$) gates are sufficient to realise a universal gate set for quantum computing \cite{Shi:2003tp}.
We implemented and performed full quantum process tomography for these two gates and for the three Pauli gates, as shown in Fig.~\ref{fig:fig2}(k), finding an average process fidelity of $\mathcal{F}_p=0.992\pm0.008$.

Note, in multi-photon experiments deviations from unit fidelity are primarily caused by imperfections in the photon source such as reduced quantum interference between different pair creation events \cite{Tanida:2012eh}, and higher order terms in the SPDC process.  
To omit these effects and measure the performance of our LPU directly we used one and two-photon ensembles to recover the implemented unitary $U^{\prime}$ \cite{Laing:2012uw} and calculated the unitary fidelity $\mathcal{F_{U}}=\abs{\text{Tr}(U^{\dag}.U^{\prime})/6}^{2}$ with the intended unitary $U$. 
For the Bell state generator, the heralded- and unheralded-CNOT respectively we found $\mathcal{F_{U}} = 0.998 \pm 0.002,0.994 \pm 0.008$ and $0.988 \pm 0.005$.\\

\begin{figure*}[h!]
\includegraphics[trim=0 0 0 0, clip, width=1.0\linewidth]{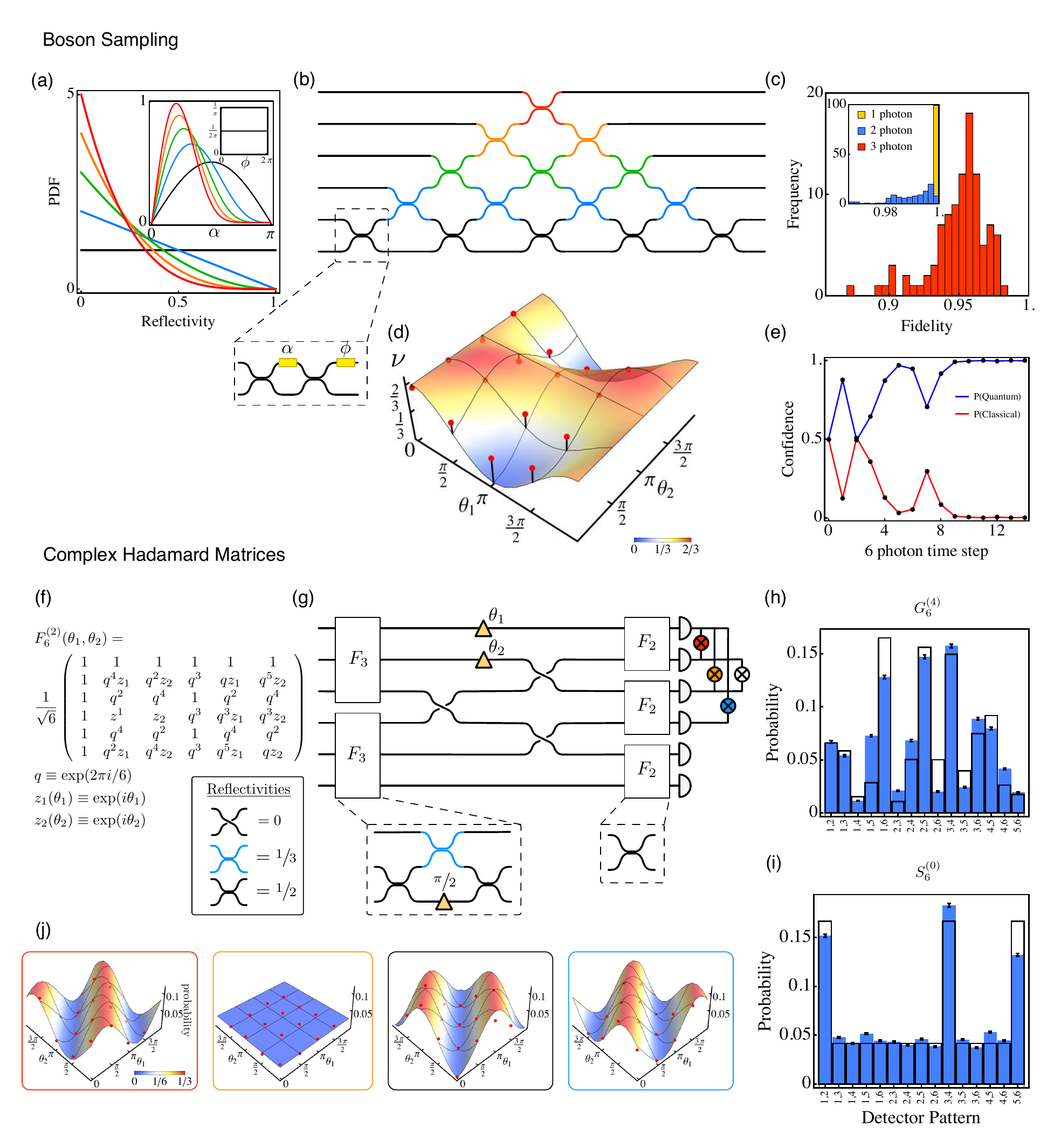}
\vspace{-1.0cm}
\caption{
Boson sampling and Complex Hadamard Matrices.
(a)
A Haar random unitary can be implemented by choosing beamsplitter reflectivities (or equivalently MZI phases $\alpha$ (inset 1)), and phase shifter values $\phi$ (inset 2) from the probability density functions (pdfs) corresponding to linear optical elements in (b).
(c)
A histogram of experimentally determined statistical fidelities for 100 three-photon boson sampling experiments, with one and two photon histograms inset.
(d)
Experimental points (red) showing measured three-photon violations ($\nu$) of the ZTL  whilst scanning across the two-parameter $F^{(2)}_{6}(\theta_1,\theta_2)$ matrices.
Minima occur at the Fourier transform point ($F_{6}$) as predicted by the theoretical manifold, black lines to guide the eye.
(e)
After 15 steps, dynamic updating with Bayesian model comparison determines with a confidence $>99\%$ that six-photon detections (from the input state $\ket{3_1,3_2}$) are sampled from an indistinguishable (quantum) as opposed to a distinguishable (classical) distribution.
(f,g)
The two parameter family of 6-dimensional CHMs $F^{(2)}_{6}$ can be directly realised on optical modes, however when mapped to our general purpose circuit, changes to these phases ($\theta_1,\theta_2$) require all on-chip phase shifters to be changed.
(h,i) 
Two-photon data for a single example of the newly discovered 4-parameter family $G^{(4)}_{6}$, which is not a member of $F^{(2)}_{6}$, and  the isolated $S^{(0)}_{6}$ case, with ideal theoretical data overlaid (error bars assume Poissonian counting statistics).
(j) 
Two photon correlation manifolds for $F^{(2)}_{6}$, with experimentally measured probabilities (red) as colour coded in (g).
}
\label{fig:fig3}
\end{figure*}

\noindent\textbf{\textit{Implementing boson sampling:}}
The realisation of a large-scale quantum computer that demonstrates an intrinsic exponential advantage over classical machines would be in conflict with a foundational tenet in computer science: the extended Church Turing thesis (ECT).  The ECT conjectures that all realistic physical systems can be efficiently simulated with a probabilistic Turing machine, or classical computer.
Designed for LO, with no requirement for quantum logic gates or qubit entangling operations, \emph{boson sampling} \cite{Aaronson:2011tja,Broome:2013ti,Spring:2013to,Crespi:2012fu,Tillmann:2012ux} is a quantum protocol that has been developed as a rapid route to challenge the ECT and demonstrate that quantum physics can be harnessed to provide fundamentally new and non-classical computational capabilities.

Based on the foundations of computer science, boson sampling is a mathematical proof (using plausible conjectures) that a many-photon state, when acted on by a large LO circuit set to implement a Haar-random unitary, will give rise to a probability distribution that cannot be efficiently sampled by a classical algorithm.
Quantum interference among the photons \cite{Hong:1987vi} contributes to the pattern of the probability distribution.
The classical intractability arises because the probability amplitude for each correlated photon detection event is given by a classically hard function, known as the \emph{permanent} \cite{Valiant:1979dl}, of the submatrix that describes a particular route of photons through the circuit.
Experimentally, each detection event represents a sample drawn from that classically forbidden probability distribution.

Acting on three-photon ensembles we programmed our device to implement 100 different boson sampling routines.  Each circuit configuration was chosen randomly from the Haar measure, which we implemented via a direct parameterisation of our circuit parameters \cite{Russell:DEP4Zrn0}, the probability density functions of which are displayed in Fig.~\ref{fig:fig3}(a).  For each implementation [Fig.~\ref{fig:fig3}(b)], we counted detection events for each of the 20 collision free ways in which three photons can exit the six output ports of the device, and in Fig.~\ref{fig:fig3}(c) plot a histogram of fidelities with statistics based on calculations of matrix permanents, finding a mean statistical fidelity of $\bar{\mathcal{F}}_s=0.950 \pm 0.020$.
These results show a test of the performance of our LPU over many circuit configurations randomly and unbiasedly chosen from the full space of all possible configurations.\\

\noindent\textbf{\textit{Verifying boson sampling:}}
An open and important question, particularly in light of the ECT, is how to verify that boson sampling continues to be governed by the laws of quantum mechanics when experiments reach the scale that classical computers can no longer simulate
\cite{Gogolin:2013uh, Aaronson:2013wc,Spagnolo:2014kh,Carolan:2014koa}.
Unlike certain algorithms for digital quantum computers, including Shor's factoring algorithm {\cite{Shor:1999ul}}, where the solution to a classically hard problem can be efficiently classically checked, there seems to be no analogous way to check that large-scale boson sampling is sampling from a probability distribution that arises from many-photon quantum interference.

While it is likely that boson sampling is in principle mathematically unverifiable, methods have been proposed to gather supporting or circumstantial evidence for the correct operation of the protocol \cite{Matthews:2013vc,Carolan:2014koa,Tichy:2014iz}.
The essence of these methods is to implement experiments that share basic quantum mechanical features with boson sampling, but where certain properties of the experimental output can be predicted and therefore checked.
The zero transmission law (ZTL) \cite{Tichy:2010gi} predicts that correlated photon detection for most of the exponentially growing number of configurations is strictly suppressed if the circuit is set to implement the Fourier transform (FT) on optical modes.  This is known because the structure of the FT allows these matrix permanents to be efficiently evaluated without explicit calculation.
Because large scale many-photon quantum interference is at the core of the ZTL, it has been proposed as a certificate for the capability of a device to implement boson sampling \cite{Tichy:2014iz}.

We programmed our device to implement 16 examples of the $F^{(2)}_{6}(\theta_1,\theta_2)$ two-parameter set of six dimensional matrices, including $F_{6}$ (the 6 dimensional FT), which occurs at $\theta_1,\theta_2=\pi,0$.  Using statistics from three-photon ensembles, we calculated the experimental violation of the ZTL as $\nu=N_s/N$, the ratio of the number of predicted suppressed events $N_s$ to the total number of events $N$; the results of which are plotted in Fig.~\ref{fig:fig3}(d) alongside the theoretical manifold.  The experimental points follow the shape of the manifold with the minimal violation of the ZTL $\nu_{min}=0.319\pm0.009$ occurring when $F_{6}$ is implemented.  The average ZTL violation of the nine points that are predicted to maximally violate is $\bar{\nu}_{max}=0.638 \pm 0.029$.  Crucially, this verification protocol is implemented in the same device and with the same procedure as that for the boson sampling experiments above.\\

\noindent\textbf{\textit{Six photon verification:}}
An essential requirement of boson sampling is the indistinguishability among photons.  With our device set to implement $F_{6}$, we injected the six-photon state $\ket{3_1,3_2}$ and counted six-photon statistics with an all-fibre beam-splitter between each output mode and two SPADs to give probabilistic number-resolved photon detection over a total of 12 SPADs.

We used Bayesian model comparison (see Ref.~\cite{Carolan:2014koa}) to update, in realtime, confidence that we are sampling from a (pre-calculated) \emph{quantum} probability distribution (arising from completely indistinguishable photons)
or from a \emph{classical} probability distribution (arising from completely distinguishable photons); as shown in  Fig.~\ref{fig:fig3}(e).
After collecting 15 six-fold coincidence events, we determine a confidence of $p=0.998$ that we are sampling from a quantum (not classical) distribution.\\
\noindent\textbf{\textit{Complex Hadamard operations:}}
The FT and $F^{(2)}_{6}$ are examples in the more general class of complex Hadamard matrices (CHMs), which are related to mutually unbiased bases \cite{Bengtsson:2007hb} and are of fundamental interest in quantum information theory \cite{Werner:2001im}.
CHMs are defined as $N \times N$ unitary matrices with entries of squared absolute value equal to $1/N$.  While this definition is straightforward, classification of these matrices is far from trivial and is concerned with identifying CHMs that are inequivalent up to pre and post multiplication with permutation matrices and diagonal unitaries \cite{Tadej:2006jq}.  In the $N=\{2,3,5\}$ case, all CHMs are equivalent to the respective FT matrix, while for $N=4$ there exists a one parameter equivalence class. 
Whilst a full classification of $N=6$ CHMs is unknown, it is currently conjectured that the set consists of an isolated matrix $S^{(0)}_{6}$ which does not belong to any family \cite{Tao:2003td}, and a newly discovered four-parameter generic family $G^{(4)}_6$ \cite{Szollosi:2010hs}.

In LO experimental implementations, discrimination among CHMs can be accomplished via the observation of characteristic patterns of photonic quantum interference \cite{Mattle:2004wy, Peruzzo:2011haa, Laing:2012bw, Spagnolo:2013iq}.  Up until now, these observations have been too experimentally challenging for the six-dimensional case.  Here, we reconstruct correlation manifolds of two-photon detection events by scanning over the $F^{(2)}_{6}$ matrices, displaying four (out of the 15 sets collected) in Fig.~\ref{fig:fig3}(j). We find a mean statistical fidelity of $\bar{\mathcal{F}}_{s} = 0.979 \pm 0.007$.

We then implemented an instance of $G^{(4)}_6$ (that is not contained in $F^{(2)}_{6}$) and $S^{(0)}_{6}$, and observed their predicted characteristic two-photon quantum interference patterns displayed in Figs.~\ref{fig:fig3}(h,i) respectively, finding statistical fidelities to be $\mathcal{F}_{s}=0.986\pm0.001$ and $\mathcal{F}_{s}=0.998\pm0.001$.  The intractability of calculating the permanents of certain CHMs is an interesting research line, as is the possibility of searching for new CHMs using photonic statistics.\\

\noindent\textbf{Discussion}\\
\noindent Photonic approaches to quantum information science and technology promise new scientific discoveries and new applications. Linear optical circuits lie at the heart of all of these protocols, and a single LPU device with the ability to arbitrarily `dial-up' such operations promises to replace a multitude of existing and future prototype systems. In the near term, combining LPUs with existing higher efficiency sources and detectors will expand their capabilities, while in the mid-term, integration of these components \cite{Silverstone:2013fu,Sprengers:2011er} with larger low-loss circuits \cite{Sohma:2006fr} will open up new avenues of research and application. \\ 

\noindent\textbf{Methods: Circuit implementation \& analysis}\\
LO protocols for quantum information processing are typically conceived and explained as a network of optical modes connected with beamsplitters and phase shifters.
Provided the total number of modes in such a protocol is not greater than that of an $m$-mode LPU,
the network can be mapped to the device via its unitary matrix description, or (due to symmetry and structure) a direct implementation can be recognised.  In the following sections we explain the circuits implemented, and note that both compiling methods are used.
While the generality of this LPU makes it well suited to implementing random unitaries,
it is not optimised for the highly structured unitaries that typically implement qubit operations.  Nevertheless, this system demonstrates fidelities greater than those reported for optical chips that have been custom made for LO quantum processing tasks.
Throughout this manuscript two error analysis methods are used: those for individual fidelities $\mathcal{F}$ calculated by propagating Poissonian count rate errors, and those for mean fidelities $\bar{\mathcal{F}}$ calculated as $1\sigma_{\mathcal{F}}$.  The bar symbol will used to denote which method is used.\\

\noindent\textbf{\textit{Heralding entangled states:}}
\nopagebreak
Our compact Bell state generator (BSG) scheme shown in Fig.~\ref{fig:fig2}(b), accepts four photons into modes $\{1,3,4,5\}$.
Detection in ancilla modes $\{3,4\}$ occurs with probability $p=2/27$ and heralds the state
$\ket{\Phi^+}=(\ket{1_{1}, 0_{2}, 1_{5}, 0_{6}}+\ket{0_{1}, 1_{2}, 0_{5}, 1_{6}}) /\sqrt{2} $ (where $\ket{n_i}$ represents $n$ photons in the $i^\text{th}$ mode)
over the remaining modes with unit probability in the ideal case.
Operating the BSG with the non-ideal state from our SPDC source requires the Bell state to be detected to remove the unwanted photon terms
(see Appendix for details).

After implementing the BSG, measurements on both qubits in the computational basis $\sigma_z \otimes \sigma_z$ are straightforward and arbitrary single qubit measurements on the $b_{0,1}$ modes $\sigma_z \otimes \sigma_{\hat{n}}$ are possible with an unused MZI.
We were also able to measure both qubits in the conjugate basis $\sigma_{\tilde{x}} \otimes \sigma_x$ by adding a fibre-looped Sagnac interferometer on the $a_{0,1}$ modes to send this qubit back through the chip to meet an MZI at the input facet 
(see Appendix for details). 

We found the statistical fidelity for the common basis measurements $\sigma_z \otimes \sigma_z$ and $\sigma_{\tilde{x}} \otimes \sigma_x$ to be $\mathcal{F}_s=0.891\pm0.015$ and $0.979 \pm 0.003$ respectively, and the uncommon bases $\sigma_z \otimes \sigma_x$ and $\sigma_z \otimes \sigma_y$ to be $0.999 \pm 0.002$ and $1.000 \pm 0.002$.\\

\noindent\textbf{\textit{Heralded quantum logic gate:}}
The heralded CNOT gate shown in Fig.~\ref{fig:fig2}(e), is a four-photon scheme \cite{Okamoto:2011ei} that has been adapted for implementation with the six-mode LPU, plus a seventh auxiliary coupling mode (mode 0) also contained on our chip.  The function of an 8th lossy balancing mode is implemented in post-processing as an equivalent reduction in detection efficiency on the lower ancilla mode (see Appendix for details).
 
Control and target qubits $c_{0,1}$ and $t_{0,1}$ are implemented via the modes $\{0,2,3,4\}$ respectively, and heralding modes via $\{1,5\}$.  In the ideal case, the detection of the two ancilla photons, which occurs with probability $p=1/16$, guarantees the CNOT logic has been performed on the two photonic qubits.  Once again, the non-ideal input state requires a computational detection to remove unwanted terms.\\

\noindent\textbf{\textit{Full quantum process tomography for 1 and 2 qubits:}}
The unheralded (or post-selected) CNOT gate shown in Fig.~\ref{fig:fig2}(h) is a two-photon scheme that requires the detection of the two photonic qubits in a computational state to signal the success of the gate with probability $p=1/9$ \cite{Ralph:2002id,Hofmann:2002hl,OBrien:2003kz}.  As the scheme requires only two photons (which our source produces with high fidelity) it allows for a less obfuscated test of our device for logical operations.

Computational states $c_{0,1}$ and $t_{0,1}$ are implemented via modes $\{2,3,4,5\}$, and MZIs (shown in orange in Fig.~\ref{fig:fig2}(h)) allow arbitrary product state preparation and measurement of both qubits, so truth tables may be taken in any bases.  
We perform full quantum process tomography using an over complete set of 324 truth tables from all possible combinations of Pauli eigenvector input states and Pauli basis measurements (see Appendix for details), finding a mean statistical fidelity between all experimentally reconstructed truth tables and ideal values to be $\bar{\mathcal{F}_s}=0.986 \pm 0.001$.

We performed single qubit process tomography on the Hadamard ($H$), $\pi/8$ ($T$) and $\sigma_x,\sigma_y,\sigma_z$ gates, finding process fidelities of
$0.995 \pm 0.001$, $0.978 \pm 0.001$, $0.997 \pm 0.001$, $0.993 \pm 0.001$, $0.994\pm0.001$ respectively,
demonstrating very high fidelities for operations requiring only one-photon interference.\\

\noindent\textbf{\textit{Haar random unitaries:}}
Along with three-photon ensembles injected into modes $\{1,2,3\}$, as a benchmarking exercise we injected the one and two photon states $\ket{1_1}$ and $\ket{1_1,1_2}$ yielding fidelities of $0.999 \pm 0.001$ and $0.990 \pm 0007$ respectively.  For three of our three-photon boson sampling unitaries (those that gave the highest, lowest, and mode fidelity) we recalculated the statistical fidelities based on the recovered unitary $U'$.  This increased the mean statistical fidelity and reduced range $\mathcal{F}^\text{max}_s-\mathcal{F}^\text{min}_s$ from $0.102$ to $0.023$; implying dialling errors are non-uniform across the space of unitaries.\\

\noindent\textbf{\textit{Compiling $F^{(2)}_{6}$:}}
Figure~\ref{fig:fig3}(g) shows a circuit customised to implement the $F^{(2)}_{6}$ family of CHMs, with the defining parameters $\theta_1,\theta_2$ explicitly realised.
However due to the particular decomposition of our device, compiling $F^{(2)}_{6}$ onto our LPU requires unique phase shift values across the entire device for each implementation $\theta_1,\theta_2\in [0,2\pi) $.

%%%%%%%% BIBLIOGRAPHY BEGINS %%%%%
% (fold)
% (end)
%%%%%%%% BIBLIOGRAPHY ENDS %%%%%

\clearpage
\newpage

\appendix

% (fold)

\section{UNIVERSAL LINEAR OPTICAL PROCESSOR} % (fold)
\label{sec:circuit}

\subsection{Photon Source} % (fold)
\label{sub:photon_source}
As shown in Fig.~\ref{fig:fig1}(b), 808nm laser light from a 140fs pulsed Titanium:Sapphire laser was attenuated via a half wave plate (HWP) and polarising beam splitter (PBS), before being frequency doubled via a type-I Barium borate (BBO) nonlinear crystal.  The subsequent 404nm light was reflected from dichroic mirrors (DM), filtering the 808nm light into a beam dump (BD), and focused onto a type-I Bismuth Borate (BiBO) nonlinear crystal to generate multiple pairs of photons through spontaneous parametric down conversion (SPDC).  After passing through an interference filter (IF) photons are collected into four polarisation maintaining fibres (PMF) and delivered to the LPU.\\
% subsection photon_source (end)

\subsection{Fabrication}
The silica-based planar waveguide circuit is fabricated on a silicon wafer using flame hydrolysis deposition, photolithography and reactive ion etching.  The germanium ($\text{GeO}_2$) doped silica core has a cross sectional dimension  of $3.5\mu m \times 3.5\mu m $, and is surrounded by a silica cladding yielding a core-cladding refractive index difference of $0.45\%$. The input and output waveguides are connected to polarization-maintaining fibre arrays by using ultraviolet curing epoxy adhesive with in- and out-coupling losses of approximately 0.4dB/facet ($9\%$).

The circuit consists of 30 tunable phase shifters, each phase of which is controlled by thin-film thermo-optic heaters fabricated on top of the circuit. The electric power required for a $2\pi$ phase shift is $\sim 0.8W$.  Directional couplers are designed to have a length of $500\mu m$ and a waveguide gap of $2\mu m$, with each directional couplers loss estimated to be less than 0.02dB ($0.4\%$).  The mean insertion loss (fibre to fibre) when averaged over all modes is 2.4dB ($42\%$).

\subsection{Electrical Control}
We developed a custom electronic circuit to deliver power to each of the on-chip heaters independently. This circuit consisted of four sets of the following: one micro-controller for communication, arithmetic, and measurement; eight 12-bit digital-to-analogue converters; eight high-power non-inverting drive amplifiers; eight current-measurement amplifiers; and eight 12-bit analogue-to-digital converters for reading out the current sourced to each heater. Each of the 32 drive ports delivered up to 20~V, with 4.9~mV resolution, and could source up to 100~mA of current, with a measurement resolution of 24~$\mu$A.

\subsection{Circuit Characterisation}
Each thermo-optic phase shifter is described by a non-linear phase-voltage relationship 
\begin{equation}
\Phi(V) = \alpha  + \beta V^2 + \gamma V^3,
\label{eq:pv}
\end{equation}
where $\Phi$ is the resulting phase shift due to an applied voltage $V$ and $\alpha$, $\beta$ and $\gamma$ are real numbers, which in general vary for each heater due to imperfections in the fabrication process.
To determine $\Phi (V)$ we inject heralded single photons into the fundamental transverse-magnetic (TM) mode of an isolated MZI ($\alpha_{i,j}$) or phase shifter ($\phi_{i,j}$) then sweep the voltage $1.8 \rightarrow 10V$ producing an interference fringe given by
\begin{equation}
C = A - B\cdot\cos\Phi(V),
\label{eq:int_fit}
\end{equation}
where $C$  is the count rate in the cross mode and $A$ and $B$, along with $\alpha, \beta$ and $\gamma$ are free parameters to be determined via a least squares fitting procedure.

The triangular arrangement of our device allows each phase shifter to be isolated and characterised independently of any other.  We first characterise the diagonal $\alpha_{1,j}$ (defined in Figure~\ref{fig:fig2} ) by:
\begin{enumerate}
	\item injecting single photons into mode 1
	\item measure interference fringe on $\alpha_{1,1}$
	\item set $\alpha_{1,1}=2\pi$
	\item repeat for $\alpha_{1,j}$ for $j=[2,5]$
\end{enumerate}
completing the characterisation of $\alpha_{1,j}$.  We repeat this procedure for $\alpha_{i,j}$, $i=[2,5]$ by injecting photons into mode $i$ and setting all $\alpha_{i-1,j}=\pi$, thus completing the characterisation of all $\alpha_{i,j}$.  We then characterise $\phi_{i,5}$ by:
\begin{enumerate}
	\item injecting single photons into mode 5
	\item set $\alpha_{i,5}=\pi$ for $j=[3,5]$
	\item set $\alpha_{i,4}=\pi$ for $j=[1,4]$
	\item set $\alpha_{2,5},\alpha_{1,5}=\pi/2$
	\item measure interference fringe on $\phi_{1,5}$
	\item repeat for $\phi_{i,5}$, with $\alpha_{i,5},\alpha_{i+1,5}=\pi/2$ and $i=[2,4]$
\end{enumerate}
completing the characterisation of $\phi_{i,5}$.  We then repeat this procedure for $\phi_{i,j}$, $j=[2,4]$ by injecting photons into mode $j$ and setting $\alpha_{i,j+1},\alpha_{i,j-1}=\pi$, thus completing the characterisation of all $\phi_{i,j}$ and therefore all phase shifters.

\subsection{Cross Talk}
Electrical cross talk was observed due to certain heaters sharing a common ground.  This is corrected for in pre-processing by fully characterising all shared resistances and  calculating the set voltages $\{V'_i\}$ which yields the actual voltages $\{V_i\}$ required for a desired phase shift.  Once corrected for no evidence of cross talk, thermal or otherwise, is observed.  

A final correction is applied due to an observed reduction in quantum interference at certain points within the chip caused by polarisation rotation.  We correct for this by inserting polarising beam splitters at the output of the chip filtering out undesired polarisation states.
% subsection characterisation_procedure (end)

\subsection{Phase accuracy} % (fold)
\label{sub:phase_accuracy}
To determine the accuracy in setting a phase we run a randomized benchmarking type experiment by injecting heralded single photons into mode 1, then setting 100 different vectors $\vec{\alpha}_{1,j}=(\alpha_{1,1},\alpha_{1,2},\alpha_{1,3},\alpha_{1,4},\alpha_{1,5})$, chosen from the Haar measure, to give a mean fidelity $\mathcal{F}^{\text{exp}}$ between our experimental and theoretical distributions.  We then run a Montecarlo simulation of the experiment by applying Gaussian noise $\delta_{\phi}$ to each phase shifter and fitting this fidelity $\mathcal{F}^m$ as a function of this noise, giving $\mathcal{F}^m(\delta_{\phi})$.  We then solve $\mathcal{F}^m(\delta_{\phi})=\mathcal{F}^{\text{exp}}$ finding $\delta_{\phi}=0.035$ rad.  This value lumps together error caused by the circuit, the control electronics and the calibration procedure, so is a good metric for our effective phase accuracy.

% subsection phase_accuracy (end)

% section circuit (end)
\section{QUANTUM PROCESS TOMOGRAPHY}
All quantum process tomography was performed as follows. Measurement effects were of the form $\ket{\psi_n}\bra{\psi_n} \otimes \ket{\psi_m}\bra{\psi_m}$ and preparations were of the form $\ket{\psi_n} \otimes \ket{\psi_m}$ where $\ket{\psi} \in \{ \ket{0},\ket{1},\ket{+},\ket{-},\ket{+i},\ket{-i} \}$. An estimate of the Choi-Jamiolkowski state $\rho_{\mathcal{E}}^{est}$ corresponding to the process was found using maximum-likelihood estimation (MLE) by minimisation of an approximation to the negative loglikelihood function:
\begin{eqnarray*}
\rho_\mathcal{E}^{est} &=& \argmin_{\rho_\mathcal{E}} \Bigg\{ \sum_{i,j}{\frac{N_{ij}(p_{ij} - d\Tr{[\rho_\mathcal{E}E_{ij}]})^2}{p_{ij}(1-p_{ij})} }\Bigg\} \\
&&\text{subject to:  } \rho_\mathcal{E}\geq 0,\text{ } \Tr_A{(\rho_\mathcal{E})}=\mathbb{1}/d 
\label{eq:minimisation}
\end{eqnarray*}

where $p_{ij}= n_{ij}/N_{ij}$ are the experimental frequencies and $E_{ij} = (\rho_i)^{\mathsf{T}}\otimes \Pi_j$, $i$ labelling the preparation and $j$ the measurement effect. To avoid problems associated with zero probabilities in MLE we use hedged MLE \cite{BlumeKohout:2010jfa}, where the experimental probabilities are adjusted to 
$$ p_{ij} = \frac{n_{ij} + \beta}{N_{ij}+K \beta} $$
where $K$ is the number of outcomes in the measurement of $p_{ij}$ and $\beta = 0.1$ is the hedging parameter. The minimisation is performed by phrasing it as a Semidefinite Program (SDP) and solved using the CSDP solver \cite{Borchers:1999jb}.

The average gate fidelity, $\mathcal{F}_{g}$, defined as $\int{d\psi \bra{\psi} U^{\dag}\mathcal{E}(\proj{\psi})U\ket{\psi}}$ was calculated from the process fidelity via the relation $\mathcal{F}_{g} = (d \mathcal{F}_p + 1)/(d+1)$

All errors were calculated by a bootstrapping method, where the raw experimental counts were resampled 100 times with Poissonian noise and the statistics of the resulting distributions analysed.

\section{QUANTUM LOGIC GATES} % (fold)
\label{sec:quantum_logic_gates}
\subsection{Unheralded CNOT}
The two photon, unheralded CNOT gate is implemented using modes \{2,3\} and \{4,5\} as dual-rail qubits, inputting photons in modes \{2,4\} and post-selecting on coincidence events in modes \{\{2,4\},\{2,5\},\{3,4\},\{3,5\}\}. To perform the gate we use the following chip phases:
$$R = 
\resizebox{0.9\columnwidth}{!}{
$
\begin{pmatrix}
(0.000, 1.231) && (\phi_{12}, \alpha_{12}) && (0.000, 3.142) && (\phi_{14}, \alpha_{14}) && (0.000, 3.142) \\
&& (0.000, \alpha_{22}) && (\phi_{23}, 1.231) && (0.000, 1.571) && (3.142, 3.142) \\
&&&& (0.000, 3.142) && (3.142, 1.571) && (0.000, 1.231) \\
&&&&&& (0.000, \alpha_{44}) && (\phi_{45},3.142) \\
&&&&&&&& (0.000, 3.142)
\end{pmatrix}
$
} $$
Where $R_{ij} = ( \phi_{ij} , \alpha_{ij} )$.
$\{\alpha_{22},\phi_{23}\}$ and  $\{\alpha_{44},\phi_{45}\}$ can be set to prepare arbitrary pure states of qubits 1 and 2 respectively, and $\{\phi_{12},\alpha_{12}\}$ and  $\{\phi_{14},\alpha_{14}\}$ can be set to perform arbitrary projective measurements on qubits 1 and 2 respectively.
To prepare Pauli eigenstates and measure Pauli observables, the following phases are set: 

\begin{center}
\begin{tabular}{c   c   c | c   c   c}
			State & $\alpha$ & $\phi$ & Observable & $\phi$ & $\alpha$ \\
			\hline
			$\ket{0}$ & $\pi$ & 0 & $\sigma_z$ & 0 & $\pi$ \\
		  $\ket{1}$ & 0 & 0 &$\sigma_x$ & 0 &  $\frac{\pi}{2} $\\
			$\ket{+}$ & $\frac{\pi}{2}$ & 0 & $\sigma_y$ & $\frac{\pi}{2} $& $\frac{\pi}{2}$ \\
			$\ket{-}$ &$\frac{\pi}{2}$ & $\pi$ &&& \\
			$\ket{+i}$ & $\frac{\pi}{2}$ & $\frac{\pi}{2}$ &&& \\
			$\ket{-i}$ &$\frac{\pi}{2} $& $\frac{3\pi}{2}$ &&& \\
		\end{tabular}
		\end{center}

\subsection{Heralded CNOT} % (fold)
\label{sub:klm_cnot}
The unitary operation (up to input and output phases) required on our six mode circuit to perform the heralded CNOT gate is given by
\begin{equation*}
U= 
\resizebox{0.9\columnwidth}{!}{
$
\begin{pmatrix}
0.476 & -0.622 & -0.440 & 0.440 & 0 & 0 \\
-0.622 & -0.476 & 0 & 0 & 0.622 & 0 \\
-0.383 & 0 & 0.293 & 0.707 & -0.383 & -0.348 \\
0.383 & 0 & 0.707 & 0.293 & 0.383 & 0.348 \\
0 & 0.622 & -0.440 & 0.440 & 0.476 & 0 \\
0.306 &0 & 0.166 & -0.166 & 0.306 & -0.870 
\end{pmatrix}
$
}\end{equation*}

Corresponding to a set of phases
$$R = 
\resizebox{0.9\columnwidth}{!}{
$
\begin{pmatrix}
(0.000, 0.992) && (4.712, 1.571) && (4.544, 4.957) && (5.375, 1.792) && (3.816, 0.000) \\
&& (0.000, 1.571) && (1.571, 0.992) && (2.188, 0.000) && (4.712, 0.000) \\
&&&& (0.000, 1.571) && (5.498, 0.000) && (1.571, 2.226) \\
&&&&&& (0.000, 3.142) && (3.142, 0.000) \\
&&&&&&&& (0.000, 0.000)
\end{pmatrix}
$
} $$
The truth table of the gate was experimentally tested by inputting four photons, two in the ancilla modes and two in the computational modes. Our photon source is based upon two independent SPDC events which produce the state $\sum_{i}\eta^i\ket{i,i}$, such that in the four photon subspace we have $\ket{\psi_4} \approx \ket{1111}+\ket{2200}+\ket{0022}$. Although we do not herald the ideal input $\ket{1111}$, the other terms should not produce output states in the computational subspace and so could be removed in measurement for this proof-of-principle demonstration.
% subsection klm_cnot (end)
% section circuit_computation (end)
\subsection{Bell state generator} % (fold)
\label{sec:heralded_entanglement_generation}
The circuit used for Bell state generation was given by the unitary matrix:

$$U = 
\resizebox{0.9\columnwidth}{!}{
$
\begin{pmatrix}
0.707 & 0.707 & 0 & 0 & 0 & 0 \\
0.408 & -0.408 & -0.577 & 0.577 & 0 & 0 \\
0.408 & -0.408 & 0.289 + 0.289i & -0.289 + 0.289i & -0.408i & -0.408i \\
0.408 & -0.408 & 0.289 - 0.289i & -0.289 - 0.289i & 0.408i & 0.408i \\
0 & 0 & 0.333 - 0.471i & 0.333 - 0.471i & 0.236 - 0.333i & 0.236 - 0.333i \\
0 & 0 & 0 & 0 & 0.707 & 0.707 
\end{pmatrix}
$
}
$$
Corresponding to a set of phases
$$R = \resizebox{0.9\columnwidth}{!}{
$
\begin{pmatrix}
(0.000, 1.571) && (0.000, 1.231) && (0.000,1.571) && (0.000, 3.141) && (0.000, 3.142) \\
&& (0.000, 3.142) && (0.000, 1.571) && (1.571, 1.231) && (0.000, 3.142) \\
&&&& (0.000, 3.142) && (0.000, 3.142) && (0.000, 1.571) \\
&&&&&& (0.000, 3.142) && (0.000, 3.142) \\
&&&&&&&& (0.000, 3.142)
\end{pmatrix}
$
}
 $$
To obtain statistics for the $\sigma_x$ measurement on qubit 1, we create a fibre looped Sagnac interferometer between the output of modes $a_0$ and $a_1$ (see Fig. 2b) injecting the photon back into the chip with a stable phase. Due to the configuration of the circuit, the photon sees a  $1/3$ reflectivity MZI between modes 2 and 3, before a $1/2$ reflectivity MZI between modes 1 and 2. This circuit implements a single qubit POVM with three elements, $\{E_+,E_-,E_1\}$ where
\begin{eqnarray*}
 E_+ &=& \frac{1}{6}\proj{1} + \frac{1}{2}\proj{0} + \frac{1}{2\sqrt{3}}\big(\ketbra{1}{0}+\ketbra{0}{1}\big) \\
 E_- &=& \frac{1}{6}\proj{1} + \frac{1}{2}\proj{0} - \frac{1}{2\sqrt{3}}\big(\ketbra{1}{0}+\ketbra{0}{1}\big) \\
 E_1 &=& \frac{2}{3}\proj{1}
\end{eqnarray*}
corresponding to detection in modes $\{1,2,3\}$  respectively. Since we cannot detect in mode 3 (this is one of the inputs), we postselect on events where the photon is detected in modes 1 and 2, which should occur with probabilities given by

\begin{eqnarray*}
p_+ &=& \frac{\Tr{(E_+ \rho)}}{ \Tr{(E_+ \rho)}+\Tr{(E_- \rho)}} \\
p_- &=& \frac{\Tr{(E_- \rho)}}{ \Tr{(E_+ \rho)}+\Tr{(E_- \rho)}} \\
\end{eqnarray*}
which leads to an expectation value
$$ p_+ - p_- = \langle \sigma_{\tilde{x}} \rangle = \frac{\frac{1}{\sqrt{3}} \langle \sigma_x \rangle }{\frac{1}{3}\Tr{(\rho \proj{1})} + \Tr{(\rho \proj{0})}} $$
such that, with knowledge of the data from the $\sigma_z$ measurement, we are able to calculate the expectation of $\sigma_x$. 

Since the experiment would require both input and detection in mode 1, we implement it in two stages, first inputting in mode 1 and detecting four-fold coincidences in modes \{2,3,4,5\} and \{2,3,4,6\}, then inputting in mode 2 and detecting four-fold coincidences in modes \{1,3,4,5\} and \{1,3,4,6\} (note that by inputting in mode 2, the state would pick up a relative $\pi$ phase shift so this is offset by setting $\phi_{35} = \frac{3\pi}{2}$). The two data sets are then normalised with respect to each other by the ratio of total counts in each experiment.

As in the case of the heralded CNOT gate, it was possible to remove events caused by the non-ideal input from the measurement statistics.

\section{COMPLEX HADAMARD MATRICES}
\subsection{Zero-Transmission Law}
The most general form of the ZTL applies to any periodic input state to the FT matrix \cite{Tichy:KcsHHzb0}. For an $n$ photon, $m$-periodic initial state $\vec{r}$ input into the FT, final states $\vec{s}$ are suppressed when the sum of the mode assignment list $\vec{d}(\vec{s})$ (i.e. the list of positions of the photons in the output state) multiplied by $m$ is not divisible by $n$
$$ \text{mod}\Big(m \sum_{i} d_{i}(\vec{s}),n\Big) \neq 0 \implies \text{Prob}(\vec{s}) = 0 $$
For our implementation, we input the periodic state $\vec{r} = \{1,0,1,0,1,0\}$ into the FT, and measure three-fold coincidences in the collision-free subspace. In this subspace, 12 out of the total of 20 outputs are suppressed and $\nu$ is calculated as the ratio of events in these 12 outputs to the total events recorded.

\subsection{$S_6^{(0)}$ \& $G_6^{(4)}$}
The isolated CHM whose output distribution is labelled $S_6^{(0)}$ in Fig.~3(i) is given by

$$S_6^{(0)}=\frac{1}{\sqrt{6}}\left(
\begin{array}{cccccc}
 1 & 1 & 1 & 1 & 1 & 1 \\
 1 & 1 & w & w & w^2 & w^2 \\
 1 & w & 1 & w^2 & w^2 & w \\
 1 & w & w^2 & 1 & w & w^2 \\
 1 & w^2 & w^2 & w & 1 & w \\
 1 & w^2 & w & w^2 & w & 1 \\
\end{array}
\right)$$
Where $w = \text{exp}(2\pi i / 3)$.

The CHM from the generic four-parameter whose distribution is labelled $G_6^{(4)}$ in Fig.~3(h), is given by
$$G_6^{(4)}=
\resizebox{0.95\columnwidth}{!}{$
\left(
\begin{array}{cccccc}
 0.408 & 0.408 & 0.408 & 0.408 & 0.408 & 0.408 \\
 0.408 & -0.085+0.399 i & -0.373-0.165 i & -0.389-0.124 i & 0.137\, -0.385 i & 0.303\, +0.274 i \\
 0.408 & -0.121-0.39 i & 0.407\, +0.028 i & -0.196+0.358 i & -0.252-0.321 i & -0.247+0.325 i \\
 0.408 & -0.333-0.236 i & -0.408+0.021 i & 0.076\, +0.401 i & 0.349\, +0.212 i & -0.093-0.398 i \\
 0.408 & -0.269+0.308 i & 0.37\, +0.173 i & -0.198-0.357 i & -0.293+0.285 i & -0.019-0.408 i \\
 0.408 & 0.4\, -0.081 i & -0.404-0.057 i & 0.298\, -0.279 i & -0.35+0.21 i & -0.352+0.206 i \\
\end{array}
\right)
$} $$

% subsection two_qubit_measurements (end)
% subsection heralded_bell_state_generation (end)
% section measurement_based_quantum_computation (end)

% (end)

\end{document}